\documentclass[secnumarabic,amssymb,nobibnotes,aps,prl,preprint,letterpaper,preprintnumbers,amsmath,superscriptaddress,floatfix,notitlepage]{revtex4-1}
\usepackage[normalem]{ulem}
\usepackage{wrapfig}
\usepackage{amsmath}
\usepackage{enumitem}
\usepackage{multirow}
\usepackage[letterpaper,margin={1in,1in}]{geometry}
\usepackage{graphicx}
\usepackage{epstopdf}
\usepackage{lineno}
\usepackage{appendix}
\usepackage{longtable}
\usepackage{array}
\usepackage{siunitx}
\usepackage[dvipsnames]{xcolor}
\usepackage{lineno}
\setcitestyle{super}
\usepackage{multirow}
\DeclareSIUnit\mrad{\milli\rad}

\usepackage{soul}
\usepackage{colortbl}
\usepackage{xcolor}
\usepackage{notes2bib}
\usepackage{listings}
\usepackage{marvosym}
\usepackage{polski}
\usepackage[UKenglish]{babel}
\usepackage[utf8]{inputenc}

\usepackage[colorinlistoftodos,prependcaption,textsize=tiny]{todonotes}

\begin{document}

\setlength{\LTcapwidth}{6.5in}

\title{Comment on newly found Charge Density Waves in infinite layer Nickelates}

\author{Jonathan Pelliciari $^+$\Letter}
\email{pelliciari@bnl.gov}
\affiliation{National Synchrotron Light Source II, Brookhaven National Laboratory, Upton, New York 11973, USA}

\author{Nazir Khan $^+$}
\affiliation{Department of Physics and Astronomy, Rutgers University, Piscataway, NJ 08854, USA}

\author{Patryk W{\k a}sik}
\affiliation{National Synchrotron Light Source II, Brookhaven National Laboratory, Upton, New York 11973, USA}

\author{Andi Barbour}
\affiliation{National Synchrotron Light Source II, Brookhaven National Laboratory, Upton, New York 11973, USA}

\author{Yueying Li}
\affiliation{National Laboratory of Solid State Microstructures, Jiangsu Key Laboratory of Artificial Functional Materials, College of Engineering and Applied Sciences, Nanjing University, Nanjing, China}
\affiliation{Collaborative Innovation Center of Advanced Microstructures, Nanjing University, Nanjing, China}

\author{Yuefeng Nie}
\affiliation{National Laboratory of Solid State Microstructures, Jiangsu Key Laboratory of Artificial Functional Materials, College of Engineering and Applied Sciences, Nanjing University, Nanjing, China}
\affiliation{Collaborative Innovation Center of Advanced Microstructures, Nanjing University, Nanjing, China}

\author{John M. Tranquada}
\affiliation{Condensed Matter and Structure of Materials Department, Brookhaven National Laboratory, Upton, New York 11973, USA}

\author{Valentina Bisogni}
\affiliation{National Synchrotron Light Source II, Brookhaven National Laboratory, Upton, New York 11973, USA}

\author{Claudio Mazzoli \Letter}
\email{cmazzoli@bnl.gov}
\affiliation{National Synchrotron Light Source II, Brookhaven National Laboratory, Upton, New York 11973, USA}

\date{\today}
\maketitle

\textbf{
Recent works ~\cite{tam_charge_2022,rossi_broken_2022,krieger_charge_2022} reported evidence for charge density waves (CDWs) in infinite layer nickelates (112 structure) based on resonant diffraction at the Ni $L_3$ edge measured at fixed scattering angle.  We have found that a measurement with fixed momentum transfer, rather than scattering angle, does not show a resonance effect. We have also observed that a nearby structural Bragg peak from the substrate appears due to third harmonic content of the incident beam, and spreads intensity down to the region of the attributed CDW order. This was further confirmed by testing a bare substrate. We suggest procedures to confirm an effective resonant enhancement of a diffraction peak.
}

\newpage

CDWs are instabilities in which electronic charge condenses in specific patterns across crystallographic structures of susceptible compounds. As such, their description (propagation vector of a scalar order parameter) strongly depends on details of the band filling and on the condensation mechanism (nesting or correlation-based, for example). Regardless of their origin, CDWs represent a great opportunity to investigate the competition among electronic states in materials. In this context, the case of layered cuprates ($3d^9$ CuO$_2$ planes) is of paramount importance for two main reasons: (i) CDWs are realized with in-plane propagation vector along a Cu-O bond direction across virtually all cuprates family, and (ii) the interaction between CDWs and high temperature superconductivity (HTSC) is highlighted by the doping evolution of the cuprates superconducting dome in their phase diagrams.
In isostructural nickelates (Ruddlesden-Popper series with $3d^8$ NiO$_2$ planes), CDWs appear with in-plane propagation vectors oriented at 45$^\circ$ to the Ni-O bond direction, while no superconductivity at sizeable temperature has been reported. 
For these reasons, the recent claim of bond-parallel CDWs in superconducting infinite layer nickelates RENiO$_2$ (RE = La, Nd) grown on (001) SrTiO$_3$ (Ni2D-STO) has spurred a lot of interest in the scientific community.

We present a critical investigation of the scattering signal reported at $\sim0$~eV energy transfer by Resonant Inelastic X-ray Scattering (RIXS), and of its attribution to CDWs in Ni2D-STO. We mostly refer to Tam \textit{et al.}~\cite{tam_charge_2022} for its completeness and extensive characterization, enabling the replication of the experimental results.

In Ni2D-STO samples the newly discovered elastic signal is described by an in-plane propagation vector of ($\frac{1}{3}$, 0), while the out-of-plane dependence remained unclear (Ref.~\onlinecite{tam_charge_2022} Fig. 3) with a maximum at the wave vector ${\bf Q}_{\rm peak}=(\frac13,0,0.31)$, in reciprocal lattice units (rlu) of the nickelate film.
The energy dependence of the diffraction intensity at ${\bf Q}_{\rm peak}$ was measured by scanning the incident photon energy $E$ while keeping the scattering angle $2\theta$ fixed (Ref.~\onlinecite{tam_charge_2022} Fig. 1~\textbf{c-e}). The diffraction signal was attributed to a CDW on the basis of its energy resonance around the Ni $L_3$ edge.
In Ref.~\cite{tam_charge_2022} the energy dependence of the diffraction intensity at ${\bf Q}_{\rm peak}$ was measured by scanning the incident photon energy $E$ while keeping the scattering angle $2\theta$ fixed (Ref.~\onlinecite{tam_charge_2022} Fig. 1~\textbf{c-e}). In this scan the momentum transfer ${\bf Q}$ necessarily moved with respect to the reciprocal lattice (RL) as $E$ varied, given that $Q=(4\pi/hc)E\sin(2\theta/2)$, where $h$ is Planck's constant, and $c$ is the speed of light. We will refer to this scan mode as E$_{{\rm fix}2\theta}$. The sample angles were adjusted to keep the in-plane component of ${\bf Q}$ fixed while scanning $E$, leaving only its out-of-plane $Q_z$ component to vary. However, to keep ${\bf Q}$ completely fixed with respect to the RL while scanning $E$, it is also necessary to adjust $2\theta$ at each incident energy as dictated by the previous equation. We refer to the latter scan mode as E$_{{\rm fix}Q}$.

\begin{figure*}[h]
\centering
\includegraphics[scale=1.0]{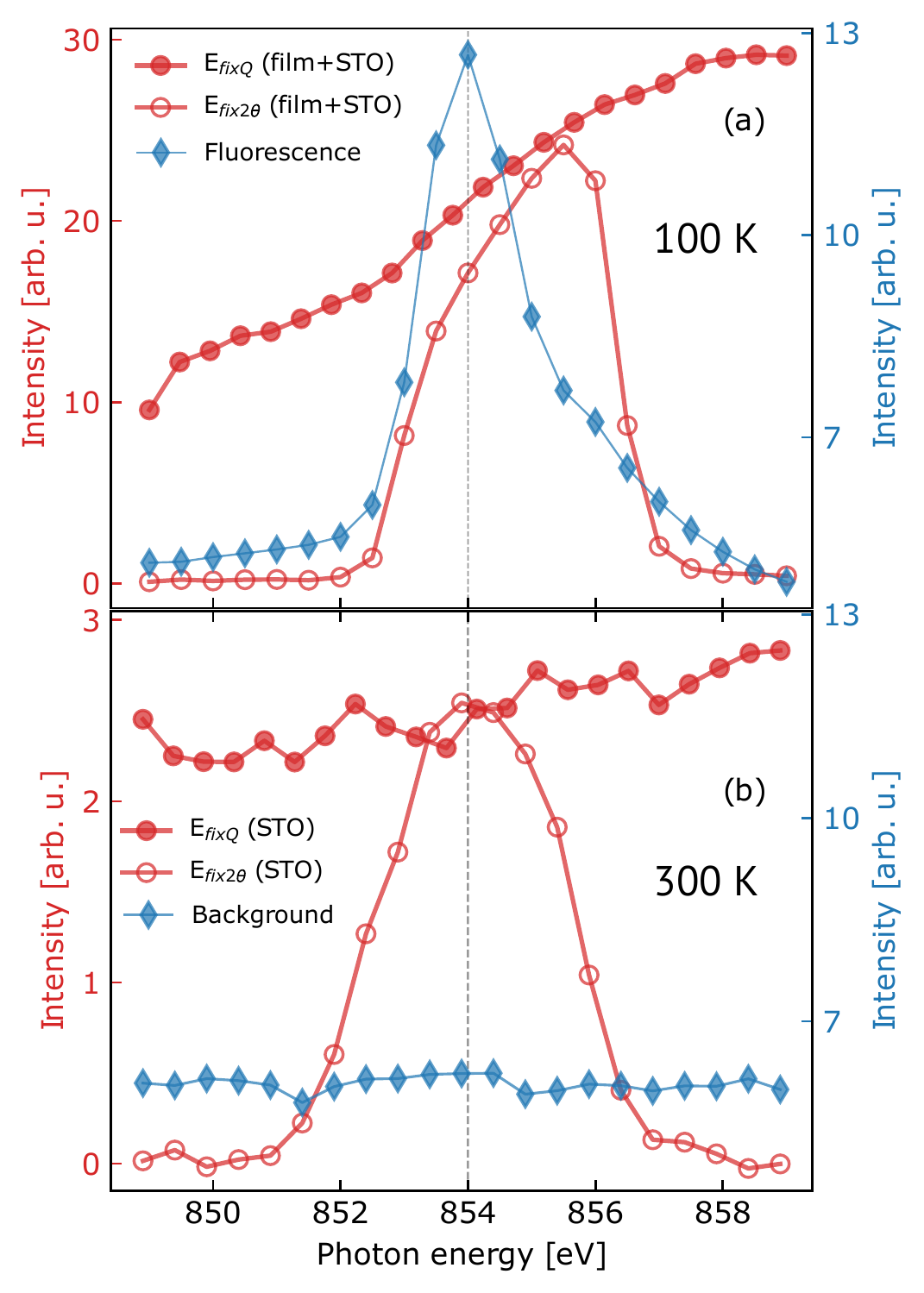}
\caption{(a) REXS E$_{fixQ}$ (filled red circles) and E$_{{\rm fix}2\theta}$ (empty red circles) collected at 100 K, on NdNiO$_2$-STO sample around the Ni $L_3$ edge. The central energy (vertical dashed line) always corresponds to ${\bf Q}_{\rm peak}=(\frac13,0,0.31)$ in film rlu, as for RIXS measurements. Both scans were background subtracted to remove a trivial fluorescence contribution (blue diamonds). (b) Same as above, but collected on a bare STO substrate with same orientation, and at 300 K.}
\label{fig1}
\end{figure*}

On a parent Ni2D-STO, equivalent to NNO$_2$-1, we measured the energy profile of ${\bf Q}_{\rm peak}$ by Resonant Elastic X-ray Scattering (REXS), see Methods. Figure~\ref{fig1}(a) reports the sample fluorescence, and the background subtracted line profiles of the two energy scans introduced above (contrary to RIXS, our REXS measurements are energy integrated on the outgoing beam, so the fluorescent background had to be removed from our data).
Our E$_{{\rm fix}2\theta}$ shows a resonant behavior with negligible off resonant contributions, roughly centered around the Ni $L_3$ edge (see the fluorescence signal provided as reference, equivalent to Ref.~\onlinecite{tam_charge_2022} Fig.~1b, $\sigma$ polarization case), making our data consistent with published results \cite{tam_charge_2022,rossi_broken_2022,krieger_charge_2022}. On the contrary, our E$_{{\rm fix}Q}$ scan displays a continuously growing signal across the scanned energy range, with no peak . The existence of a signal at all energies in E$_{{\rm fix}Q}$ proves that its dominant character is of non-resonant nature, where its intensity variation along the energy spanned can be ascribed to the geometrical characteristics of the sample (thin film on a substrate).
As such, it is evident that the difference between the two energy scans here reported stems from the different ways in which the detector intercepts surrounding elastic scattering with non-resonant behavior.
Moreover, by surfing the RL we discovered that our measured responses are dominated by proximity to scattering associated with the $(1,0,1)$ allowed reflection of the SrTiO$_3$ (STO) substrate. This peak cannot be reached at energies around Ni $L_3$ (x-ray wavelength $\lambda = hc/E$), but it results from $\lambda/3$ contamination.  Thus, interpreting the STO Bragg peak as if it were measured at $\lambda$ corresponds to a peak at ${\bf Q}_{\rm STO} = (\frac{1}{3},0,\frac{1}{3})$ in STO rlu. In terms of indexing, the NdNiO$_2$ film is epitaxial with STO; therefore, the in-plane wave vector remains the same, while a difference in the out-of-plane lattice parameters leads to an apparent incommensurate peak position along the $L$ index.  Taking into account those differences leads to an estimate of ${\bf Q}_{\rm STO} = (\frac13,0,0.28)$ in film rlu, close to ${\bf Q}_{\rm peak}$.
It is worth noticing that changes in Sr doping and temperature modify the $c$ axis lattice parameter of NdNiO$_2$ film, and consequently the RL position of ${\bf Q}_{\rm STO}$ in film rlu \cite{li_superconducting_2020,zeng_phase_2020}.
In turn, this can cause an induced intensity dependence measured at ${\bf Q}_{\rm peak}$, as it shifts with respect to ${\bf Q}_{\rm STO}$.

Our evidence on ${\bf Q}_{\rm peak}$ in Ni2D-STO question the resonant nature of the signal reported by RIXS. So, to further test its origin, we investigated a bare STO substrate in the same conditions as for Ni2D-STO. Measurements were performed at room temperature (RT), where the elastic signal is reported to be minimum in RIXS experiments (Ref.~\onlinecite{tam_charge_2022}, Fig. 4\textbf{c} top panel).
Figure~\ref{fig1}(b) shows our results. In STO no Ni ions are present to generate fluorescence, and an essentially constant diffuse background is detected [red trace in Fig.\ref{fig1}(b)]. Nonetheless, background subtracted energy scans performed at ${\bf Q}_{\rm peak}$ on the bare substrate show an apparent resonance across Ni $L_3$ if acquired by E$_{{\rm fix}2\theta}$, and an essentially constant diffracted intensity by E$_{{\rm fix}Q}$.
Some noticeable differences with the film remain, such as energy position, line shape, and signal intensity. However, these can be ascribed to specific details, and they do not affect qualitatively our observations.

In conclusion, data acquired at ${\bf Q}_{\rm peak}$ on Ni2D-STO contain a substantial non-resonant contribution, resulting from tails of a substrate Bragg peak in third harmonic.
While this does not prove the absence of CDWs in Ni2D-STO, the presence of an unexpected elastic non-resonant contamination suggests a different approach to distinguish spurious signals from proper electronic correlations in RIXS measurements is needed as done by REXS measurements in the past~\cite{ghiringhelli_long-range_2012}. We propose E$_{{\rm fix}Q}$ scans at different $Q$ locations as the new standard for the determination of resonant contributions in quasi-elastic features (the so called REXS-in-RIXS). This is essential in the case of 3D structures but it proves helpful in 2D systems as well, as in this case.

\newpage

\section{Author contributions}
$^+$ These authors contributed equally to this work.

\section{Methods}
\paragraph{Growth and sample preparation}
The NdNiO$_3$ samples were grown on the TiO$_2$-terminated SrTiO$_3$ (001) substrates by molecular beam epitaxy using a DCA R450 MBE system. The NdNiO$_3$ film was grown at 600$^{\circ}$ C and under an oxidant background pressure of $\approx$4.0$\times$10$^{-6}$ Torr (distilled ozone). The sample was sealed in a vacuum chamber together with $\approx$0.1 g CaH$_2$ powder, and then heated to 280$^{\circ}$ for 4h, with warming (cooling) rate of 10-15$^{\circ}$ C / min to attain the infinite-layer phase. More details could be found in previous report \cite{li_impact_2021}.

The NdNiO$_2$film was 18 unit cells thick. The lattice parameter of the NdNiO$_2$ are $a = b =3.905$~\AA~and $c = 3.353$~\AA.

\paragraph{X-ray scattering measurements}
The scattering experiment was performed on the TARDIS endstation of 23-ID-1, NSLS-II, BNL. The x-ray polarization was kept in $\sigma$ channel. The sample was at 100K by a LHe flow cryostat.

\section{Acknowledgements} 
This research used resources (2-ID and 23-ID-1) of the National Synchrotron Light Source II, a U.S. Department of Energy (DOE) Office of Science User Facility operated for the DOE Office of Science by Brookhaven National Laboratory under Contract No. DE-SC0012704. This investigation was supported by the Laboratory Directed Research and Development project of Brookhaven National Laboratory No. 19-013 and 21-037. This work was supported by the U.S. Department of Energy (DOE) Office of Science, Early Career Research Program. N.K. was supported by the National Science Foundation, Grant No. DMR-2103625. Y.L. and Y.N. would like to acknowledge support from National Key R\&D Program of China (Grants No. 2021YFA1400400 and 2022YFA1402502). J.M.T. was supported at Brookhaven National Laboratory by DOE's Office of Basic Energy Sciences, Division of Materials Sciences and Engineering. 

\section{Competing Interests}
The authors declare no competing interests.

\section{Data availability}
\vspace{-0.2cm}
Relevant data are available upon reasonable request from the corresponding authors.

\end{document}